\documentclass[twocolumn,aps,prl,showpacs, amsmath,amssymb, superscriptaddress]{revtex4}

\usepackage{graphicx}
\usepackage{dcolumn}
\usepackage{bm}

\begin{document}

\preprint{APS/123-QED}

\title{Robust Quantum State Transfer in Random Unpolarized Spin Chains}

\author{Norman Y. Yao}
\affiliation{Department of Physics, Harvard University, Cambridge, MA 02138, U.S.A.}
\author{Liang Jiang}
\affiliation{IQI, California Institute of Technology, Pasadena, CA 91125, U.S.A.}
\author{Alexey V. Gorshkov}
\affiliation{Department of Physics, Harvard University, Cambridge, MA 02138, U.S.A.}
\affiliation{IQI, California Institute of Technology, Pasadena, CA 91125, U.S.A.}
\author{Zhexuan Gong}
\affiliation{Department of Physics and MCTP, University of Michigan, Ann Arbor, MI 48109, U.S.A.}
\author{Alex Zhai}
\affiliation{Department of Mathematics, Harvard University, Cambridge, MA 02138, U.S.A.}
\author{Luming Duan}
\affiliation{Department of Physics and MCTP, University of Michigan, Ann Arbor, MI 48109, U.S.A.}
\author{Mikhail D. Lukin}
\affiliation{Department of Physics, Harvard University, Cambridge, MA 02138, U.S.A.}

\date{\today}
\begin{abstract}

We propose and analyze a new approach for quantum state transfer between remote spin qubits. Specifically, we demonstrate that coherent quantum coupling between remote qubits can be achieved via certain classes of random, unpolarized (infinite temperature) spin chains. Our method is robust to coupling strength disorder and does not require manipulation or control over individual spins. In principle, it can be used to attain perfect state transfer over arbitrarily long range via purely Hamiltonian evolution and may be particularly applicable in a solid-state quantum information processor.   As an example, we demonstrate that it can be used to attain strong coherent coupling between Nitrogen-Vacancy centers separated by micrometer distances at room temperature. Realistic imperfections and decoherence effects are analyzed.  

\end{abstract}

\maketitle

In addition to diverse applications ranging from quantum key distribution to quantum teleportation \cite{Bennett93, Gisin02}, reliable quantum state transfer between distant qubits forms an essential ingredient of any scalable quantum information processor~\cite{Lloyd93}. However, most direct qubit interactions are short-range and the corresponding interaction strength decays rapidly with physical separation. For this reason, most of the feasible approaches that have been proposed for quantum computation rely upon the use of quantum channels which serve to connect remote qubits; such channels include: electrons in semiconductors ~\cite{Greentree04}, optical photons~\cite{Blinov04, Duan04, Moehring07, Togan10}, and the physical transport of trapped ions~\cite{Kielpinski02}. Coupled quantum spin chains have also been extensively studied~\cite{Bose03, Petrosyan10, Christandl04, Burgarth07, Yung05, Difranco08, Kay07, Feldman10, Clark05, Venuti07, Gualdi08, Paternostro05, Tsomokos07, Wojcik05, Banchi10}. A key advantage of such spin chain quantum channels is the ability to manipulate, transfer, and process quantum information utilizing the same fundamental hardware \cite{Gong07}; indeed, both quantum memory and quantum state transfer can be achieved in coupled spin chain arrays \cite{RTNVQC}, eliminating the requirement for an external interface between the quantum channel and the quantum register. Prior work on spin chain quantum channels has focused on three distinct regimes, in which the spin chain is either initialized~\cite{Bose03, Feldman10, Venuti07, Gualdi08, Banchi10}, engineered~\cite{Christandl04, Malinovsky97, Feldman09} or dynamically controlled~\cite{Karbach05, Cappellaro07, Zhang09, Fitzsimons06, Yung05}. 

An important application of spin-chain mediated coherent coupling is in the context of realizing a room temperature quantum information processor based upon localized spins in the solid-state~\cite{Stoneham09}. In this case, it is difficult to envision mechanical qubit transport, while other coupling mechanisms are often not available or impose additional prohibitive requirements such as cryogenic cooling~\cite{Togan10}. At the same time, long spin chains are generally difficult to polarize, impossible to control with single-spin resolution, and suffer from imperfect spin-positioning \cite{Paternostro05, Tsomokos07}; such imperfections can cause both on-site and coupling disorder, resulting in localization~\cite{Evers08}. For these reasons, a detailed understanding of quantum coherence and state transfer in random spin chains with a limited degree of external control is of both fundamental and practical importance. 

\begin{figure}
\includegraphics[width=3.4in]{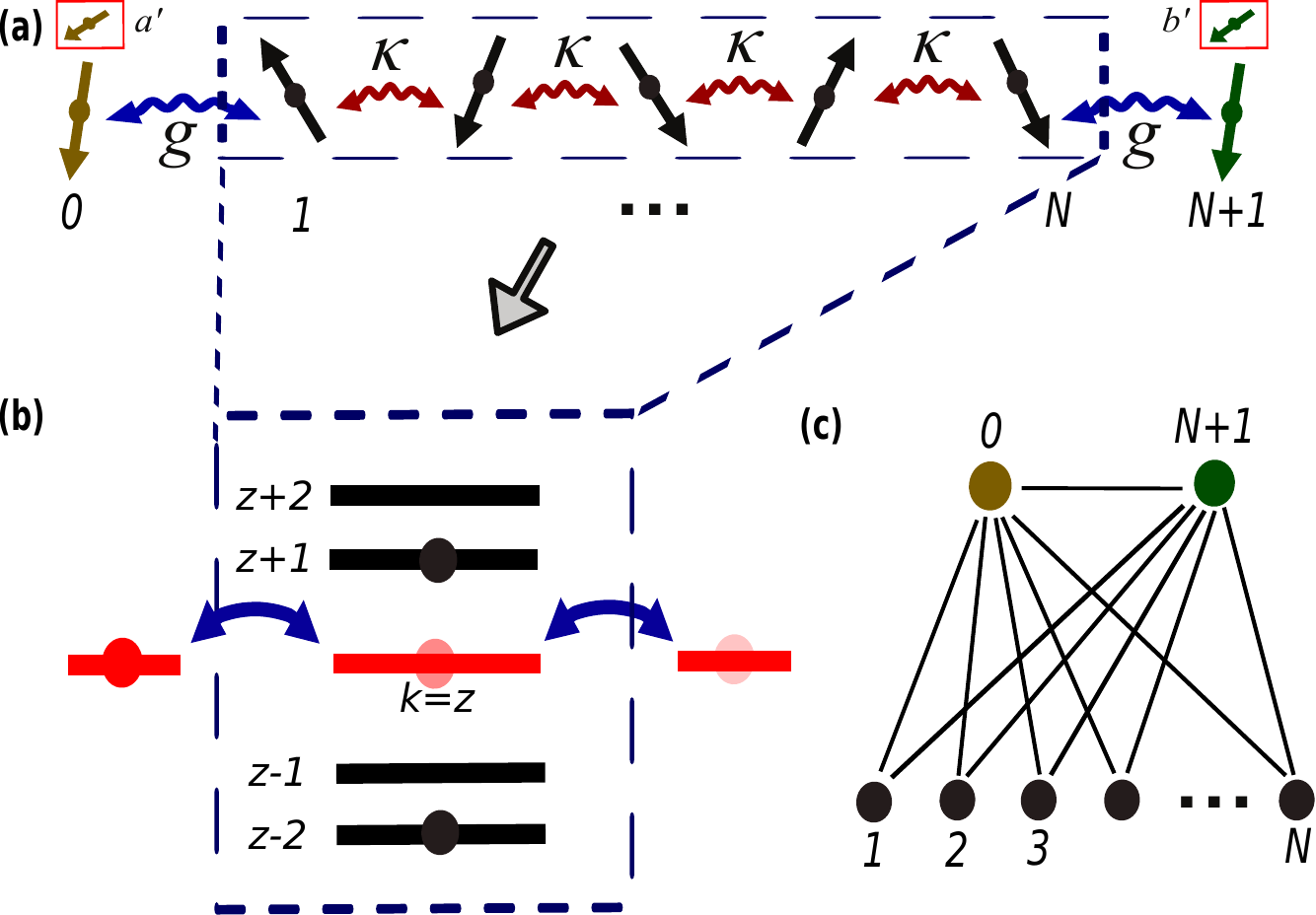}
\caption{\label{fig:Fig1} (color online). (a) Distant spin qubits coupled by an unpolarized spin-chain quantum channel with $g$, the coupling between qubits (yellow, green) and the spin chain and $\kappa$, the coupling between intra-chain elements. The spin chain can be re-expressed in terms of free fermions via the Jordan-Wigner transformation, wherein the hopping strength is characterized by $\kappa$. Boxed spins, labeled $a'$ and $b'$, represent additional spin qubits that can correspond to the memory of a quantum register or ancillary qubits associated with quantum information encoding. (b) By ensuring that the end spins are resonant with a single fermion mode ($k=z$), unpolarized spin-chain state transfer becomes analogous to fermionic tunneling. Maintaining $g \ll \kappa / \sqrt{N}$ ensures that off-resonant coupling to other fermionic modes can be neglected and enables state transfer independent of the intermediate spin-chain state. (c) Graph-like state generated by FFST, between the qubits and the intermediate spin chain~\cite{Clark05}. Each line represents a controlled-phase gate.}
\end{figure}

In this Letter, we propose and analyze a novel method for quantum state transfer (QST) in an unpolarized, infinite temperature spin chain. In contrast to prior work, the method requires neither external modulation of the Hamiltonian evolution nor spin chain engineering and initialization. Furthermore, it is robust to specific, practically important types of disorder.  The key idea of our approach is illustrated in Fig.~\ref{fig:Fig1}(a). The two spin qubits at the ends of the spin chain can be initialized and fully controlled, while the coupling between these remote qubits is mediated by a set of intermediate spins, which can not be initialized, individually controlled, or optically detected.  We assume that the qubit-chain coupling $g$, which can be variably adjusted, and the intrachain coupling $\kappa$, which is fixed, are characterized by short-range XX interactions. The essence of the state transfer is the long-range coherent interaction between the spin qubits, mediated by a specific collective eigenmode of the intermediate spin chain. This mode is best understood via Jordan-Wigner (JW) fermionization, which allows for the states of an XX spin chain to be mapped into the states of a set of non-interacting spinless fermions. In this representation, the state transfer is achieved by free fermion tunneling, as shown in Fig.~\ref{fig:Fig1}(b). In what follows, we show that the initial state of the intermediate chain does not affect the tunneling rate associated with free fermion state transfer (FFST), allowing for the implementation of a SWAP operation between the end qubits after a period of unitary evolution.  
 
To be specific, we consider an XX Hamiltonian governing two distant qubits connected by a quantum channel consisting of a spin-$1/2$ chain 
\begin{equation}
H = H_{0} + H' 
\end{equation}
\noindent with $H_{0}=\sum_{i=1}^{N-1} \kappa (S_{i}^{+} S_{i+1}^{-} + S_{i}^{-} S_{i+1}^{+})$ and $H'= g(S_{0}^{+} S_{1}^{-} + S_{N+1}^{+} S_{N}^{-} + \mbox{ h.c.})$, as shown in Fig.~\ref{fig:Fig1}(a). Here, $S^{\pm} = S^{x} \pm i S^{y}$, where $\vec{S} = \vec{\sigma}/2$ and $\vec{\sigma}$ are Pauli spin operators ($\hbar = 1$). We consider the limit $g \ll \kappa$, and work perturbatively in $H'$. Upon introducing fermi operators $c_{i} = e^{i \pi \sum_{0}^{i-1} S_{j}^{+} S_{j}^{-}} S_{i}^{-}$,  $H_{0}$ is transformed to $H_{0} = \sum_{i=1}^{N-1} \kappa (c_{i}^{\dagger} c_{i+1} + c_{i} c_{i+1}^{\dagger})$, wherein conservation of total spin z-projection becomes conservation of fermion number~\cite{Lieb61}. The subsequent diagonalization of this tight-binding Hamiltonian occurs through an orthogonal transformation $f_{k}^{\dagger} = \frac{1}{A} \sum_{j=1}^{N} \sin \frac{j k \pi}{N+1} c_{j}^{\dagger}$ with $k = 1, \cdots , N$ and $A = (\frac{N+1}{2})^{1/2}$, yielding $H_{0} = \sum_{k=1}^{N} E_{k} f_{k}^{\dagger} f_{k}$, where $E_{k} = 2 \kappa \cos  \frac{k \pi}{N+1}$~\cite{Lieb61}. The perturbation Hamiltonian is likewise transformed to
\begin{equation}
H' = \sum_{k=1}^{N} t_{k} (c_{0}^{\dagger} f_{k} +  (-1)^{k-1}c_{N+1}^{\dagger} f_{k} + \mbox{ h.c.}),
\end{equation}
\noindent where $t_{k}=\frac{g}{A} \sin \frac{k \pi}{N+1}$. We begin by restricting our discussion to odd $N$, where there exists a single zero energy fermionic mode in the intermediate chain corresponding to $k=z\equiv(N+1)/2$. Thus, the two end spins are resonantly coupled to the zero energy fermion by $H'$, and under the assumption that the tunneling rate $t_{z}\sim g/A$ is much smaller than the fermion detuning, $|E_{z}-E_{z\pm1}|\sim \kappa/N$, off-resonant coupling to other fermionic modes can be neglected. Upon absorbing a phase factor of $(-1)^{z-1}$ into $c_{N+1}^{\dagger}$, evolution is governed by the effective Hamiltonian, $H_{eff} = t_{z} (c_{0}^{\dagger} f_{z} + c_{N+1}^{\dagger} f_{z} + \mbox{ h.c.})$, which describes resonant fermionic tunneling, as shown in Fig.~\ref{fig:Fig1}(b).

Unitary evolution under $H_{eff}$ for a time $\tau = \frac{ \pi}{ \sqrt{2} t_{z}}$ results in $U_{eff} = e^{-i \tau H_{eff}} = (-1)^{f_{z}^{\dagger} f_{z}} (1 - (c_{0}^{\dagger} + c_{N+1}^{\dagger})(c_{0}+ c_{N+1}))$. Upon projection to the subspace spanned by $\{(1, c_{0}^{\dagger}, c_{N+1}^{\dagger}, c_{0}^{\dagger}c_{N+1}^{\dagger} ) |00 \rangle_{0,N+1}\}$, the effective evolution can be expressed as 
\begin{equation}
U_{eff}^{fermi} = (-1)^{n_{0}+n_{N+1}+n_{z}} (-1)^{n_{0}n_{N+1}} \mbox{SWAP}_{0,N+1},
\end{equation}
\noindent where $n_{\theta} = f_{\theta}^{\dagger} f_{\theta}$ is the fermion number operator. Hence, as desired, time evolution under $H_{eff}$ swaps the quantum state of the two end fermions. However, in addition to the SWAP gate and single fermion rotations, the end fermions are entangled through a controlled phase gate $\mbox{CP}_{0,N+1}=(-1)^{n_{0}n_{N+1}}$, which arises from fermionic anticommutation relations~\cite{Gorshkov10, Clark05, Yung05}. Before discussing this entanglement, let us first consider the analogous prescription in the spin basis. 

We consider a generic initial state $\Phi_{i} = (\alpha | \downarrow \rangle + \beta | \uparrow \rangle)_{0} \otimes (\alpha' | \downarrow \rangle + \beta' | \uparrow \rangle)_{N+1} \otimes \Psi_{M,n_{z}}$ where $\Psi$ represents the intermediate spin chain state, characterized as the co-eigenstate of commuting operators $M = \sum_{j=1}^{N} S_{j}^{+} S_{j}^{-}$ and $n_{z}$. After fermionization, evolution and inversion back to the spin basis, the final spin chain state becomes
\begin{equation}
\Phi_{f} = (\prod_{j=1}^{N} \mbox{CP}_{0,j} \mbox{CP}_{N+1,j})\mbox{CP}_{0,N+1} \mbox{SWAP}_{0,N+1} \Phi_{i}
\end{equation}
\noindent up to single qubit rotations. In this basis, the Wigner-strings become controlled-phase gates and generate a graph-like entangled state between the two end spins and the intermediate spins, as shown in Fig.~\ref{fig:Fig1}(c)~\cite{Clark05}.

Despite this entanglement, the use of a simple two-qubit encoding can achieve coherent quantum state transfer~\cite{Markiewicz09}. The quantum information is encoded in two spins, $a$ and $a'$, with logical basis $|\downarrow \rangle = | \downarrow \rangle_a |\downarrow \rangle_{a'}$, $|\uparrow \rangle = | \uparrow \rangle_a |\uparrow \rangle_{a'}$. After encoding, one first performs FFST between spins $a$ and $b$ via the unpolarized spin chain, and then, repeats the operation between spins $a'$ and $b'$, as shown in Fig.~\ref{fig:Fig1}(a). Finally, the quantum information is decoded by applying a CNOT gate between spins $b$ and $b'$, after which, the information has been coherently mapped to spin $b$. Thus, we have demonstrated the ability to perform QST between spatially separated spin qubits. Furthermore, as detailed in the subsequent section on experimental realizations utilizing Nitrogen-Vacancy registers, we offer an alternative solution which achieves remote coupling of spatially separated quantum registers through a dual-transfer protocol.

\begin{figure}
\includegraphics[width=3.4in]{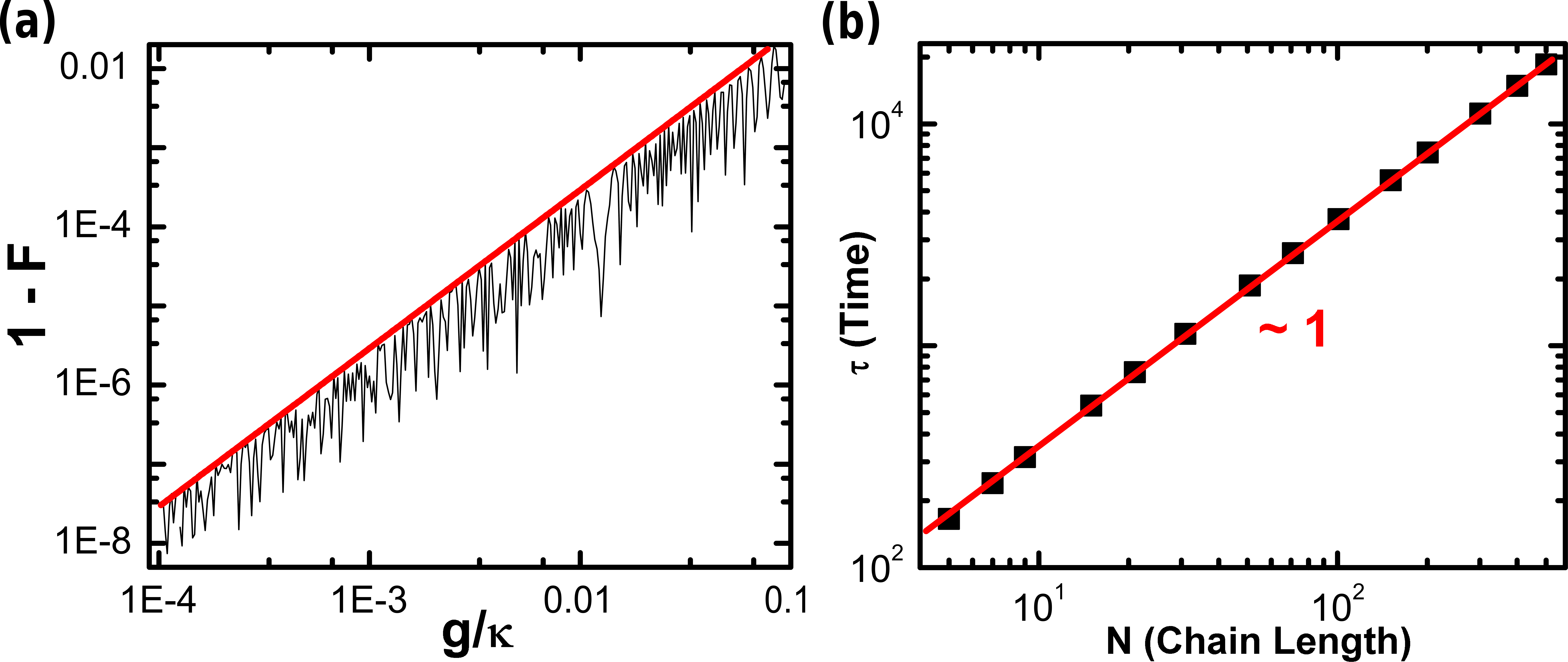}
\caption{\label{fig:Fig2} (color online).  (a) Numerical simulation of the infidelity of QST for $N=7$ as a function of $g/\kappa$ depicting fluctuations in the infidelity. The numerical infidelity is bounded by the theoretical estimate (bold line). (b) For a chosen tolerable infidelity $\epsilon_{0} = 10^{-3}$, the minimum time $\tau$ (in units of $1/\kappa$), required for state transfer scales linearly with chain length. } 
\end{figure}

To confirm perfect quantum state transfer, we perform numerics, as shown in Fig.~\ref{fig:Fig2}.  Specifically, we calculate the average fidelity,~$F =\frac{1}{2}+ \frac{1}{12} \sum_{i=1,2,3} \text{Tr} \left [ \sigma^{i} \mathcal{E} (\sigma^{i}) \right ]$,~of two-qubit encoded state transfer, where $\mathcal{E}$ represents the quantum channel consisting of encoding, state transfer and decoding~\cite{Nielsen02}. This average fidelity can be expressed in terms of elements of the matrix $e^{-iK \tau}$, where $K$ is the $N\times$$N$ coupling matrix of the full Hamiltonian found in Eq.~(1), $H = \sum_{i,j} K_{i,j} S_{i}^{+}S_{j}^{-}$ \cite{in prep}; crucially, this allows for simulations of channel fidelity in extremely long spin chains, since diagonlization of the full Hilbert space is no longer necessary.  In finite chains of fixed length, the infidelity, $\epsilon=1-F$, varies as a function of $g/\kappa$, as shown in Fig.~\ref{fig:Fig2}(a). This infidelity results from the leakage of quantum information into the off-resonant modes of the intermediate spin chain, and can be analytically expressed, in the limit $g \ll \kappa$, as $\epsilon \approx \sum_{k\neq z}\frac{5}{3} \left ( \frac{t_{k}}{E_{k}} \right )^{2}[1+(-1)^{k+z} \cos (E_{k}\tau)]$, where $z = \frac{N+1}{2}$~\cite{in prep}. In this limit, the analytic expression is in exact agreement with the numerics, and is upper-bounded by the theoretical estimate,  $\sum_{k\neq z} \frac{10}{3}\left ( \frac{t_{k}}{E_{k}} \right )^{2}$, as shown in Fig.~\ref{fig:Fig2}(a).   

Utilizing the analytic upper bound for a given chain length $N$, a given intrachain coupling $\kappa$, and a given tolerable infidelity $\epsilon_{0}$, we can compute the maximum allowed $g$ and hence the minimum state transfer time $\tau$. By contrast to direct dipole-dipole interactions, which would depict a cubic scaling of $\tau$ with $N$, the time required for FFST scales linearly with chain length, as shown in Fig.~\ref{fig:Fig2}(b)~\cite{Lieb72}. Intuitively, this results from the fact that the condition on $t_{z}$ allowing for off-resonant coupling to be neglected is $t_{z} \ll \kappa/N$, implying that $\tau \sim 1/t_{z} \sim N/\kappa$.

\emph{Extensions.}--While we have chosen to focus on the case of odd $N$ length intermediate chains, the extension to even $N$ is directly analogous. In even $N$ chains, since the fermion eigenspectrum is symmetric about $E=0$, no fermionic eigenmode is initially resonant with the end spin qubits. However, by introducing a controllable detuning to the end spins, $H_{\Delta} = \Delta  (S^{z}_{0} + S^{z}_{N+1})$, it is possible to choose an N-dependent $\Delta$ such that the end spins are resonant with any single fermion eigenmode in both even \emph{and} odd $N$ cases~\cite{RTNVQC}. In particular, for $\Delta = E_{k}$, resonant tunneling will occur at the rate $t_{k}$, allowing for control over the speed of FFST. 

We now generalize our analysis and consider optimizing the FFST protocol in the context of realistic imperfections including disorder and decoherence. On-site and coupling disorder cause localization, asymmetry of the eigenmodes, and changes in the statistics of the eigenenergies~\cite{Paternostro05, Tsomokos07, Evers08}. In the thermodynamic limit in 1D, localization occurs for any amount of disorder; thus, it will be necessary to utilize eigenmodes whose localization length is sufficiently large relative to the chain length, thereby rendering such modes effectively extended and viable for QST.  Crucially, in the case of particle-hole (PH) symmetric disorder (e.g.~coupling-strength disorder), there exists an extended critical state at $E=0$ with a diverging localization length; this ensures the existence of an extended eigenmode with a known eigenenergy, suggesting that FFST is intrinsically robust against coupling-strength disorder~\cite{Evers08}. In the case of on-site disorder, random modulation of the on-site potential may be able to restore PH symmetry \cite{Zou10}; in cases where this is insufficient, it is possible to characterize the energy spectrum and coupling strengths of the intermediate spin chain solely through tomography of a single end spin \cite{Burgarth09}. This characterization will help allow for the identification of a suitable, extended eigenmode.  

However, the existence of an extended mode is not sufficient to ensure state transfer as disorder also enhances off-resonant tunneling rates and causes the eigenmode wavefunction amplitude to become asymmetric at the two ends of the chain. Despite such imperfections, by individually tuning the qubit-chain couplings, $g_{L}$ (left) and $g_{R}$ (right), it is possible to compensate for eigenmode asymmetry; furthermore, sufficiently decreasing the magnitude of the qubit-chain coupling ensures that off-resonant tunneling can safely be neglected, even in the presence of disorder.  

In addition to disorder, decoherence of the spin qubits and the intermediate spin chain places a stringent lower bound on the values of $g_{L}$ and $g_{R}$, since $\tau \sim \sqrt{N}/g$ \cite{RTNVQC}. Thus, an interplay of disorder and decoherence will ultimately limit the experimental realization of FFST; further numerical exploration of such an interplay will provide insight into the relevant constraints~\cite{RTNVQC, in prep}. 

\emph{Experimental Realization.}--Both the necessity and realization of FFST can be evinced by considering Nitrogen-Vacancy (NV) registers in diamond, which have extremely long room-temperature coherence times \cite{RTNVQC}. In particular, the imperfect conversion of NV centers from single Nitrogen impurities results in substantial spatial separation between individual registers. However, the unconverted spin-1/2 Nitrogen impurities form a natural spin chain connecting remote registers. At ambient temperatures, the Nitrogen impurity spin chain, which is optically unaddressable, would be unpolarized and hence, the proposed scheme would be essential to enable distant NV register coupling. 
  
Thus, we envision an array of two-qubit NV registers connected by a quantum channel consisting of spin-1/2 implanted Nitrogen impurities \cite{RTNVQC}. Recent experiments have demonstrated the ability to fully manipulate the two-qubit NV register corresponding to the NV nuclear spin, which serves as the memory qubit, and the NV electronic spin, which is used to initialize, readout, and mediate coupling to the intermediate spin chain~\cite{Childress06, Balasubramanian09, Rittweger09, Maurer10}. The effective Hamiltonian described in Eq. (1) can be achieved in such a mixed spin system via dynamic decoupling~\cite{RTNVQC}, and the qubit-chain coupling $g$ can be fully tuned by utilizing the structure of the NV center ground-state manifold~\cite{RTNVQC}.  To apply arbitrary two qubit gates between the nuclear memory of distant NV registers: 1) SWAP the state of the nuclear and electronic spin of the first register 2) apply FFST between the electronic spins of the two registers 3) apply a CP-gate between the electronic and nuclear spin of the second register 4) repeat (2) and (1) to return the nuclear memory of the first register and disentangle from the intermediary chain. Together with single qubit rotations, such an implementation of FFST achieves a universal set of gates and hence computation in an array of NV registers connected by Nitrogen impurity spin chains.

In summary, we have proposed a robust method to coherently couple spatially separated quantum registers by means of an unpolarized spin chain. The proposed method is examined in the context of NV diamond centers, where its direct application can potentially allow for the realization of a scalable room-temperature quantum information processor~\cite{RTNVQC}. While we have focused on the specific case of an XX chain, the conceptual framework can be used in a wide range of systems to achieve QST through effective eigenmode tunneling. For example, QST in an unpolarized chain can also be achieved in the transverse field Ising model, where in contrast to the XX chain, the JW transformation yields a fermionic Hamiltonian which no longer conserves fermion number~\cite{in prep}. In fact, all Hamiltonians that are quadratic in bose and fermi operators can be exactly diagonalized and thus provide a natural starting point to further explore eigenmode-mediated QST. Finally, the proposed approach may also provide insight into entanglement generation in a many-body system and the dynamics of the disorder-driven localization transition.

We gratefully acknowledge conversations with G. Giedke, J. I. Cirac, G. Goldstein, T. Kitagawa, J. Maze, P. Maurer, E. Togan, Y. Chu, J. Otterbach, C. Laumann, C. Mathy, A. S{\o}rensen, J. Preskill, N. Schuch, S. F. Lichtenstein and Y. T. Siu.  This work was supported by the NSF, DOE, CUA, DARPA, AFOSR MURI, NIST, Lee A. DuBridge Foundation and the Sherman Fairchild Foundation.

\end{document}